# Electrically modulated photoluminescence in ferroelectric liquid crystal

Prasun Ganguly, D. Haranath, and A. M. Biradar*

*CSIR-National Physical Laboratory, Dr. K. S. Krishnan Road, New Delhi-110012, India*

T. Joshi,[†] and S. Singh

*Department of Physics, Banaras Hindu University, Varanasi- 221 005, India*

Electrical modulation and switching of photoluminescence (PL) have been demonstrated in pure deformed helix ferroelectric liquid crystal (DHFLC) material. The PL intensity increases and peak position shifts towards lower wavelength above a threshold voltage which continues up to a saturation voltage. This is attributed to the helix unwinding phenomenon in the DHFLC on the application of an electric field. Moreover, the PL intensity could be switched between high intensity (field-on) and low intensity (field-off) positions. These studies would add a new dimension to ferroelectric liquid crystal's application in the area of optical devices.

---

*Author to whom correspondence should be addressed. E-mail: abiradar@mail.nplindia.ernet.in

[†]Also working at CSIR-National Physical Laboratory, Dr. K. S. Krishnan Road, New Delhi-110012, India.

Recently, luminescent materials which have the ability to switch their photoluminescent properties in response to various external stimuli have attracted much attention due to their potential application for memory devices, sensors, security materials, and informational displays.[1–6] Liquid crystal (LC) is also evolved as a better candidate for stimuli responsive luminescent materials because of the dynamic response of LC molecules.[7,8] The luminescent properties of LC is known to be modulated by light irradiation,[9] electric field,[10] and temperature.[11] Electric-field controlled photoluminescence (PL) in LC has been observed in terms of tuning and switching by various workers.[12-14] Tong et al.[10] reported the first LC gels whose fibrous aggregates are strongly fluorescent and demonstrated the possibility of using an electric field to switch the PL of this new material. Wide tuning of PL in dye-doped nematic liquid crystal has been observed by Ozaki et al.[13] Likewise, the field modulated PL of LC monosubstituted polyacetylene have been reported by Huang et al.[14] However, most of these studies have been reported in nematic and cholesteric LCs by adding dyes, quantum dots, gels etc. Ferroelectric liquid crystals (FLCs), which are a special kind of LCs, have attracted a great deal of attention by global researchers due to their interesting features such as low driving voltage, fast response, non-volatile memory and wide viewing angle characteristics.[15–17] **In our group, PL studies have also been done on FLCs with the aim to improvise the photoluminescence (PL) intensity of the existing FLCs by incorporating gold nanoparticles and ZnS quantum dots (QDs).[18,19] However, electrically modulated PL study in FLC has not been reported so far.** Due to ultra-short pitch and high sensitivity to electric field,[20] deformed helix FLC (DHFLC), a member of FLC, has been chosen as a better option for such studies.

With this in view, we have studied and reported the effect of electric field on PL of a DHFLC material, namely, FLC 6304 (Rolic, Switzerland). **In the present work, modulation of PL characteristics (both the intensity and wavelength) by applying an electric field over the DHFLC material without any doping has been emphasized. In addition to this, switching of the PL intensity has also been achieved. The probable mechanism has been discussed on the basis of field-induced helix unwinding model in the DHFLC material.**

The LC sample cell for the present study was prepared using highly conducting (~30 Ω/□) indium tin oxide (ITO) coated glass plates. The desired electrode patterns on the ITO substrates were achieved using a photolithographic technique. The active electrode area was 45 mm × 45 mm. The thickness of the cell was maintained uniformly ~5 μm using Mylar spacers. The homogeneous alignment was obtained using rubbed polyimide technique. The FLC 6304 material was filled in isotropic phase by means of capillary action and then cooled gradually to room temperature. The phase sequence of this DHFLC material is as follows:

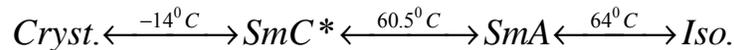

$$Cryst. \xleftrightarrow{-14^0 C} SmC^* \xleftrightarrow{60.5^0 C} SmA \xleftrightarrow{64^0 C} Iso.$$

The room temperature PL excitation and emission spectra of the filled LC sample cell was recorded in the fluorescence mode using luminescence spectrometer (Edinburgh, F900, UK) equipped with a xenon lamp. A dc regulated power supply was used for applying external electric field across the LC cell. Dielectric permittivity of the sample was measured using an impedance analyzer (Wayne Kerr, 6540 A, UK).

The PL excitation spectrum of the filled LC sample cell was recorded over the range 200-360 nm using the luminescence spectrometer. Initial parameters like slit width, excitation step, and dwell time were kept constant at 5 nm, 1 nm, and 0.1 s respectively, for the sample. It was found that the FLC 6304 material has a clear absorption peak at 333 nm **while that of ITO coated glass plates is at 258 nm**. The PL emission of this DHFLC material was then recorded by registering the excitation wavelength at 333 nm. Figure 1 shows the PL emission spectra ranging from 350-410 nm at various applied voltages (0-30 V). The emission spectrum of the DHFLC material is found to be voltage dependent. Both PL peak position and intensity get modified by changing the applied voltage. It is observed that there is almost no change in peak position when the applied voltage is less than 3 V. At around 3 V, it gets slightly shifted to lower wavelength and this shift gets pronounced as the voltage is further increased to 4 V. This shifting continues up to 10 V and remains constant thereafter. It is a well known fact that in DHFLC material the helix is easily distorted by electric field which leads to a change in the refractive index of the material.[21] This further results in shifting of the PL peak position.[22,23] Hence, the observed shifting in the peak position with voltage is attributed to the change in refractive index due to helix distortion in the DHFLC material. For better understanding, this voltage induced helix distortion is shown schematically in Fig. 2. At low voltages (< 3 V), only deformation of helix occurs [Fig. 2 (a)] where the pitch is almost same which results in no change in the PL peak position.[24] **The pitch of the present DHFLC material is 0.35 μm which is almost linearly dependent on the electric field above a threshold voltage which is defined as value of the voltage required to switch the DHFLC molecules.**[25,26] Above this threshold voltage (3 V), the twist walls become unstable and the helix starts unwinding which leads to the shift in the peak position towards lower wavelength.[20] Due to the surface inhomogeneity there is a voltage range (4 < V < 10 ) where unwound and helical parts coexist [Fig. 2 (b)],[20] which is manifested as the observed shifting in the PL emission peaks in this range. **In this intermediate voltage range, in some areas the helical structure remains stable while in other areas switching occurs between two unwound states which lead to non-uniform switching. See supplementary media files (i) at [*URL will be inserted by AIP*] for [*non-uniform electro-optical switching at 4 V and 8 V using 500 mHz ac signal*]**. At and above saturation voltage (≥10 V), which is defined as the voltage where **complete switching of DHFLC molecules takes place,** the helix is unwound everywhere [Fig. 2 (c)] and the switching becomes uniform. **See supplementary media file (ii) at [*URL will be inserted by AIP*] for [*uniform electro-optical switching at 10 V using 500 MHz ac signal*]**. This complete unwinding of the helix causes no peak shifting above 10 V.

Figure 3 shows the variation of PL intensity with the applied voltage. It is clear from the figure that the PL intensity increases above a threshold voltage (3 V) and continues up to a saturation voltage (10 V). **Figure 3 (inset) shows the response time of the material as a function of applied voltage which further confirms these values of voltages.** The mechanism behind these observations can similarly be explained on the basis of voltage-stimulated unwinding of the helical structure in the DHFLC material. It is known that upon application of voltage, the helix gets destabilized resulting in a highly light scattering state.[12] Hence, the excitation photons undergo multiple scattering by this highly scattering state before emission out of the LC cell to give the enhanced PL brightness. As discussed before, the helix unwinding process in FLC 6304 material starts from around 3 V and retains up to 10 V, which explains the observed increase in the PL intensity in this voltage range. **Another possible reason for this variation of the PL** **intensity could be due to supra-helical structure. However, such structures have not been observed in the present FLC material which is in concurrence to earlier reported work.[26] See supplementary figure file (iii) at [*URL will be inserted by AIP*] for [optical micrographs at different voltages showing the textures]**.

Apart from this, the PL intensity of the DHFLC material can be switched between low intensity (field-off) and high intensity (field-on), which is shown in Fig. 4. The change in PL intensity between these switching states has been visualized and captured as a digital image [Fig. 4 (a)]. The increase in the light intensity provides evidence of electrically switchable PL in the DHFLC material. Time-dependent switching of PL intensity in response to several cycles of field-off (0 Vµm$^{-1}$) and field-on (6 Vµm$^{-1}$) is shown in Fig. 4 (b) whereas the corresponding PL emission spectra ($\lambda_{ex}$= 333 nm) are compared in the inset of Fig. 4 (b). The time interval (2 min) between two cycles is accounted to the total time required for each PL scan and stabilization of the applied field. It can be seen that the PL intensity quickly reaches to maximum during the field-on state while it goes to minimum in field-off state. This fast switching can be understood as FLC 6304 material is reported to have fast response time (~1-5 ms) to the external applied field.[27] This electrically switchable and repeatable PL intensity reveal the possibility of using DHFLC material in optical switches.

In summary, we present the first study of using the electric field to modulate and switch the PL intensity of DHFLC material. The field-induced helix distortion in DHFLC material undoubtedly plays the key role for this remarkable behavior. This study may lay an important foundation for developing the next generation FLC-based electrically modulated optical devices.

The authors sincerely thank Prof. R. C. Budhani, Director, CSIR-National Physical Laboratory, New Delhi, for continuous encouragement and interest in this work. Authors (PG and TJ) are thankful to Council of Scientific and Industrial Research (CSIR) and University Grants Commission (UGC), respectively, for providing financial assistance.

**Figure captions**

**FIG. 1.**  (Color online) PL emission spectra of pure FLC 6304 material excited with 333 nm at various voltages.

**FIG. 2.**  (Color online) Schematic showing helix unwinding process in DHFLC material: (a) helix deformation at low voltages (0 – 3 V), (b) co-existence of unwound and helical parts (4 – 10 V), and (c) complete helix unwinding above 10 V.

**FIG. 3.**  Variation of PL intensity excited at 333 nm and **response time (inset)** with the applied voltage of FLC 6304 material.

**FIG. 4.**  (Color online) Electrical switching of PL intensity of FLC 6304 material: (a) photos showing light emission from LC sample cells under excitation (b) Time-dependent switching of PL intensity in response to several cycles of field-off (0 Vμm$^{-1}$) and field-on (6 Vμm$^{-1}$), inset shows the corresponding emission spectra.